\documentclass[11pt]{article}
\usepackage[dvips]{color}
\usepackage{epsfig}
\usepackage{amsmath}
\usepackage{graphicx}
\date{}
\begin{document}
\title{{\bf Late time acceleration in a non-commutative model of modified
cosmology}}

\author{B. Malekolkalami$^1$\thanks{%
e-mail: b.malakolkalami@uok.ac.ir},\,\, K. Atazadeh$^2$\thanks{%
e-mail: atazadeh@azaruniv.ac.ir}\,\, and B. Vakili$^3$\thanks{%
e-mail: b.vakili@iauctb.ac.ir}
%EndAName
\\\\
$^1${\small {\it Department of Physics, University of Kurdistan,
Pasdaran St., Sanandaj, Iran}}\\
$^2${\small {\it Department of Physics, Azarbaijan Shahid Madani University, 53714-161, Tabriz, Iran}}\\
$^3${\small {\it Department of Physics, Central Tehran Branch,
Islamic Azad University, Tehran, Iran}}}

\maketitle

\begin{abstract}
We investigate the effects of noncommutativity between the
position-position, position-momentum and momentum-momentum of a
phase space corresponding to a modified cosmological model. We show
that the existence of such noncommutativity results in a Moyal
Poisson algebra between the phase space variables in which the
product law between the functions is of the kind of an
$\alpha$-deformed product. We then transform the variables in such a
way that the Poisson brackets between the dynamical variables take
the form of a usual Poisson bracket but this time with a
noncommutative structure. For a power law expression for the
function of the Ricci scalar with which the action of the gravity
model is modified, the exact solutions in the commutative and
noncommutative cases are presented and compared. In terms of these
solutions we address the issue of the late time acceleration in
cosmic evolution.

\vspace{5mm}\noindent\\
PACS numbers: \vspace{0.8mm}98.80.-k, 04.20.Fy, 04.50.+h
\newline Keywords: Late time acceleration, Noncommutative modified
cosmology,
\end{abstract}

\section{Introduction}
\indent According to the astronomical evidences and corresponding
observations, we live in an accelerated expanding universe. To
explain this acceleration, many models have been proposed which in
general may be classified in two categories. The first is the models
which deal with ordinary (or dark) matter and energy including
scalar fields in the context of the Einstein's general theory of
relativity \cite{dark}. The second one is the so-called modified
theories of gravity for which there are good theoretical reasons to
consider the possibility that gravity is not described precisely by
Einstein's theory (frame) but rather by some alternative theories
\cite{Sut}.

In the past decade, investigations of the noncommutative (NC)
structures in the field of the first category have caused a renewed
interest on noncommutativity in the classical and quantum levels. In
these studies, the influence of noncommutativity (in the
configuration space or even in momentum sector) has been explored by
formulation of a version of NC cosmology in which noncommutativity
shows itself as a deformation parameter in the classical or quantum
commutation relations between minisuperspace \cite{9}-\cite{11} or
phase-space \cite{12}-\cite{plb} variables while the geometrical
structure of the underlying space-time remains unchanged.

However, as far as the authors are aware, studying the subject in
the level of the second category (modified theories of gravity) has
seldom been studied in the literature. As an excellent review in
which many aspects of gravitational issues have been studied in the
context of extended theories of gravity, we can refer to the work of
Capozziello and De Laurentis \cite{Lau}. Apart from the conceptual
problems in astrophysics and cosmology, this work has dealt with the
recently emerged problems in the subject of gravity such as
inflation, dark energy, dark matter, large scale structure, quantum
gravity and so on. In particular, it is shown in \cite{Lau} that the
space-time metric deformations can be related to the generation of
gravity theory. To do this, they considered the deformations as
extended conformal transformations to address the relations between
the Jordan and the Einstein frames. Although our main motivation in
this work is the idea of the space-time deformation such as what
mentioned in \cite{Lau}, we do not do this at the level of the
space-time geometry. Instead, our purpose in this work is to build
an NC scenario for a classical modified gravity model. To do this,
we will investigate the effects of noncommutativity either in the
configuration space with the NC parameter $\theta$ or in the
momentum sector whose NC parameter is $\beta$. Specially, we will
show that the latter case causes more ability in tuning and problem
debugging of time solutions of interested variables.

The work is organized as follows: Section 2 is devoted to introduce
the modified gravity model we are going to deal with in the rest of
the paper. In the first subsection of the third section we have
presented the time evolution of model and in its second subsection
we will solve the corresponding cosmology but this time in the NC
framework. We finally summarize the work in section 4 by a
discussion about how the cosmological picture is changed when one
takes into account the NC considerations.

\section{The model}
Let us start with an $f(\mathcal{R})$ theory whose action in the
Jordan frame may be written as
\begin{equation}\label{A1}
\mathcal{A}=\frac{1}{2\kappa}\int d^4x\sqrt{-g} f(\mathcal{R}),
\end{equation}
where $g$ is the determinant of the metric tensor $g_{\mu \nu}$,
${\cal R}$ is the Ricci scalar corresponding to $g_{\mu \nu}$ and we
let $f$ to be an arbitrary function which its form will be specified
later. Under the conformal transformation
\begin{equation}\label{A2}
\tilde{g}_{\mu\nu}=e^\phi g_{\mu\nu},
\end{equation}
where
\begin{equation}\label{A3}
\phi=\ln f'(\mathcal{R}),
\end{equation}
the above action can be expressed in the Einstein frame  as
\begin{equation}\label{A4}
\mathcal{A}=\frac{3}{2\kappa}\int
d^4x\sqrt{-\tilde{g}}\Bigl[\frac{\tilde{\mathcal{R}}}{3}-\frac{1}{2}(\tilde{\nabla}\phi)^2-\frac{V(\phi)}{3}\Bigr],
\end{equation}
in which tilde-quantities should be evaluated with respect to the
metric $\tilde{g}_{\mu\nu}$ and
\begin{equation}\label{A5}
V(\phi)=\frac{\mathcal{R}f'(\mathcal{R})-f(\mathcal{R})}{f'(\mathcal{R})^2},
\end{equation}where a prime represents differentiation with respect
to ${\cal R}$. As the background geometry we consider a spatially
flat FRW space-time where following \cite{Louko} its metric is given
by
\begin{equation}\label{A6}
ds^2=-\frac{\tilde{N}^2(t)}{\tilde{a}^2(t)}dt^2+\tilde{a}^2(t)\delta_{ij}dx^idx^j,
\end{equation}where
$\tilde{N}(t)$ and $\tilde{a}(t)$ being the lapse function and the
scale factor, respectively. The choice of the square of the scale
factor dividing the lapse function as the $00$ component of the
metric seems to be a little unusual. However, as we will see it
simplifies the calculations and gives a Hamiltonian function with a
quadratic form. By substituting (\ref{A6}) into (\ref{A4}) it is
easy to obtain an effective point-like Lagrangian for the model as

\begin{equation}\label{A8}
{\cal L}=\frac{1}{2\tilde{N}}\left(- \tilde{a}^2
\dot{\tilde{a}}^2+\frac{1}{4}\tilde{a}^4
\dot{\phi}^2\right)-\frac{\tilde{N}}{6}\tilde{a}^2V(\phi),
\end{equation}where we have re-scaled the scale factor as $\tilde{a}\longrightarrow \tilde{a}/\sqrt{2}$.
We may write the Lagrangian in a simpler form by introducing a new
set of variables \cite{Maharana}
\begin{equation}\label{A9}
x=\frac{\tilde{a}^2}{2}\cosh
\phi,\hspace{.5cm}y=\frac{\tilde{a}^2}{2}\sinh \phi,
\end{equation}
in terms of which Lagrangian (\ref{A8}) takes the form
\begin{equation}\label{B1}
{\cal
L}=\frac{1}{2\tilde{N}}\left(\dot{y}^2-\dot{x}^2\right)-\frac{\tilde{N}}{3}\left(x-y\right)e^{
\phi}V(\phi).
\end{equation}
So far, we did not consider a specific form for the function
$f(\mathcal{R})$. From now on, let us suppose that this function has
the form $f(\mathcal{R})=f_0\mathcal{R}^m$, for which according to
(\ref{A5}) the potential function will be
\begin{equation}\label{B2}
V(\phi)=\frac{m-1}{m^2f_0}\mathcal{R}^{2-m}.
\end{equation}
Also equation (\ref{A3}) gives
\begin{equation}\label{B3}
\mathcal{R}=(mf_0)^{\frac{1}{1-m}}e^{\frac{\phi}{m-1}},
\end{equation}
and finally, in terms of $\phi$ the potential becomes
\begin{equation}\label{B4}
V(\phi)=V_0e^{-\alpha\phi},
\end{equation}
where we have introduced the abbreviations
\begin{center}
$V_0=3\frac{m-1}{m}(mf_0)^{\frac{1}{1-m}}\hspace{.75cm}$ and $\hspace{.75cm}\alpha=\frac{m-2}{m-1}$.
\end{center}
For the numeric value $m=3/2$ $(\alpha=1)$ the last term in the
Lagrangian (\ref{B1}) will be simplified as
\begin{equation}\label{B6} {\cal
L}=\frac{1}{2\tilde{N}}\left(\dot{y}^2-\dot{x}^2\right)-\tilde{N}V_0\left(x-y\right),
\end{equation}
with the corresponding Hamiltonian
\begin{equation}\label{B7}
{\cal H}=\tilde{N}\Bigl(-\frac{1}{2}p_x^2+\frac{1}{2}p_y^2+V_0\left(x-y\right)\Bigr),
\end{equation}
which satisfies the constraint
\begin{equation}\label{B71}
{\cal H}\equiv 0.
\end{equation}
Therefore, we are led to a two dimensional minisuperspace model
whose coordinates vary in the intervals $0<a<\infty$,
$-\infty<\phi<+\infty$. The boundaries of this two dimensional
manifold may be divided into the nonsingular and singular types. By
definition, at the nonsingular boundary we have $a=0$ with
$|\phi|<+\infty$, while at the singular boundary, at least one of
the two variables is infinite \cite{Vil}. When we use the variables
$x$ and $y$, introduced in (\ref{A9}), to recover the above
mentioned minisuperspace, the range of them should be restricted as
$x>0$, $x>|y|$, and therefore the nonsingular boundary may be
represented by $x=y=0$. In what follows we are going to deal with
the resulting cosmological solution based on the Hamiltonian
(\ref{B7}) which seems to have more suitable form for study the
subject in the NC framework.
\section{Cosmological dynamics}
In the context of the Hamiltonian formalism the classical solutions
of the model described by Hamiltonian (\ref{B7}) can be easily
obtained. Since, we are interested to know how NC considerations
affect the usual commutative solutions, in what follows we will
discuss these two cases separately and compare the results with each
other.
\subsection{Commutative case}
In this case the Poisson brackets for the classical phase space
variables will be
\begin{equation}\label{J}
\left\{x_i,x_j\right\}=\left\{p_i,p_j\right\}=0,\hspace{.5cm}\left\{x_i,p_j\right\}=\delta_{ij},
\end{equation}
where $x_i(i=1,2)=x,y$ and $p_i(i=1,2)=p_x, p_y$. Assuming
$\tilde{N}=1$, the equations of motion become
\begin{equation}\label{K}
\dot{x}=\left\{x,{\cal
H}\right\}=-p_x,\hspace{.5cm}\dot{p_x}=\left\{p_x,{\cal
H}\right\}=-V_0,
\end{equation}
\begin{equation}\label{L}
\dot{y}=\left\{y,{\cal
H}\right\}=p_y,\hspace{.5cm}\dot{p_y}=\left\{p_y,{\cal
H}\right\}=V_0.
\end{equation}
These equations can be immediately integrated to yield the solutions
\begin{equation}\label{M}
x(t)=\frac{1}{2}V_0
t^2-p_{0x}t+x_0,\hspace{.5cm}p_x(t)=-V_0t+p_{0x},
\end{equation}
\begin{equation}\label{N}
y(t)=\frac{1}{2}V_0
t^2+p_{0y}t+y_0,\hspace{.5cm}p_y(t)=V_0t+p_{0y},\end{equation}where
$x_0$, $p_{0x}$, $y_0$ and $p_{0y}$ are some integration constants
whose values, due to the zero energy condition (\ref{B71}), are
restricted by the following relation
\begin{equation}\label{O}
p_{0x}^2-p_{0y}^2=2V_0 (y_0-x_0).
\end{equation}
We notice that the equations (\ref{M}) and (\ref{N}) have the same
form of ones for a particle moving in a plane with a constant
acceleration ${\bf a}=(V_0,V_0)$ while $-p_x(t)$ and $p_y(t)$ play
the role of its velocity components. Note that the condition $x>0$
implies that $p_{0x}^2-2V_0x_0<0$, thus, equation (\ref{O}) results
in $p_{0y}^2-2V_0y_0<0$, which means that $y>0$. Therefore, in
classical cosmology only half of the minisuperspace: $x>y>0$ or
$(\tilde{a}>0, \phi <0)$ is recovered by the dynamical variables
$x(t)$ and $y(t)$. Now, going back to the relation (\ref{A9}) we can
find the scale factor and scalar field as
\begin{equation}\label{O1}
\tilde{a}(t)=\left[8|p_{0x}|(V_0t^3+2x_0t)\right]^{1/4},
\end{equation}
\begin{equation}\label{O2}
\phi
(t)=\frac{1}{2}\ln\left(\frac{V_0t^2+2x_0}{2|p_{0x}|t}\right),
\end{equation}
in which we have set $x_0=y_0$ and $p_{0x}=p_{0y}$ compatible to
(\ref{O}). Now, let us assume the following form for the metric
tensor appeared in action (\ref{A1})

\begin{equation}
g_{\mu\nu}=\mbox{diag}(-N^2, a^2, a^2, a^2).\end{equation} Then,
from (\ref{A2}), after some simple algebra, we get the Gordan frame
counterparts of the lapse function and the scale factor as
\begin{equation}\label{O0}
N^2(t)=\frac{e^{-\phi}}{\tilde{a}^2}=\frac{1}{2(V_0t^2+2x_0)},
\end{equation}
\begin{equation}\label{O11}
a^2(t)=\tilde{a}^2e^{-\phi}=4|p_{0x}|t.
\end{equation}
As is clear, the scale factor describes a deceleration universe,
more accurately a pure radiation universe ($a(t)\propto \sqrt{t}$)
for $t> 0$. Also, the lapse function has a singularity at $t=0$ for
initial condition $x_0=0$. In the next section, we will see how
these problems may be solved in the NC frame.
\subsection{Noncommutative case}
Let us now go forward by considering the noncommutativity concepts
in classical cosmology obtained above. Classically, noncommutativity
can be described by a deformed product known as the star product law
between two arbitrary functions on a $2N$-dimensional phase space as
\begin{equation}\label{Q}
(f*_{\alpha}g)(q,p)=\exp\left[\frac{1}{2}\alpha^{ab}\partial^{(1)}_a
\partial^{(2)}_b\right]f(q_1,p_1)g(q_2,p_2)|_{q_1=q_2=q,p_1=p_2=p},
\end{equation}
such that
\begin{equation}\label{R}
\alpha_{ab}=\left(%
\begin{array}{cc}
\theta_{ij} & \delta_{ij}+\sigma_{ij} \\
-\delta_{ij}-\sigma_{ij} & \beta_{ij} \\
\end{array}%
\right),\end{equation}where $\theta$ and $\beta$ are the $N\times N$
antisymmetric matrices which represent the noncommutativity in
coordinates and momenta, respectively. The relation between the
above star-product of phase space functions and the Poisson brackets
becomes more clear if the formula (\ref{Q}) is expressed as follows
\cite{Moyal}
\begin{equation}\label{R1}
f*g=fg+\frac{1}{2}\{f,g\}+\sum_{k=2}^{\infty}
\left(\frac{1}{2}\right)^k\frac{1}{k!}D_k(f,g),
\end{equation}where the bidifferential operator $D_k$ is defined as

\begin{equation}\label{R2}
D_k(f,g)(q,p)=\left[\left(\frac{\partial}{\partial
q_1}\frac{\partial}{\partial p_2}-\frac{\partial}{\partial
q_2}\frac{\partial}{\partial p_1}\right)^k
f(q_1,p_1)g(q_2,p_2)\right]_{{q_1=q_2=q,p_1=p_2=p}}.
\end{equation}
According to (\ref{Q})-(\ref{R2}) one is led to the following
definition for the Moyal bracket as a deformed Poisson bracket
\begin{equation}\label{S}
\{f,g\}_\alpha=f*_\alpha g-g*_\alpha
f=\{f,g\}+\sum_{k=2}^{\infty}\left(\frac{1}{2}\right)^k\frac{1}{k!}\left[D_k(f,g)-D_k(g,f)\right],\end{equation}which
looks like an $\alpha$-commutation relation between two function $f$
and $g$ (a more detailed analysis of Moyal and Poisson brackets is
given in \cite{Moyal}). Now, a simple calculation shows that the
Poisson brackets between the phase space coordinates will be
deformed as
\begin{equation}\label{T}
\{x_i,x_j\}_\alpha=\theta_{ij},\hspace{.5cm}\{x_i,p_j\}_\alpha=\delta_{ij}+\sigma_{ij},\hspace{.5cm}\{p_i,p_j\}_\alpha=\beta_{ij}.
\end{equation}However, we may find a transformations on the classical phase
space, as \cite{Witten}
\begin{equation}\label{U}
x'_i=x_i-\frac{1}{2}\theta_{ij}p^j,\hspace{.5cm}p'_i=p_i+\frac{1}{2}\beta_{ij}x^j,
\end{equation}such that the canonical pair $(x_i, p_j )$ obeys the usual Poisson algebra (\ref{J}) and then
\begin{equation}\label{V}
\{x'_i,x'_j\}=\theta_{ij},\hspace{.5cm}\{x'_i,p'_j\}=\delta_{ij}+\sigma_{ij},\hspace{.5cm}\{p'_i,p'_j\}=\beta_{ij},
\end{equation}where $\sigma_{ij}=-\frac{1}{8}\left(\theta_i^k\beta_{kj}+\beta_i^k\theta_{kj}\right)$.
These commutative relations have the same form as (\ref{T}) with
this difference that here the brackets are the ordinary Poisson
brackets. Consequently, for introducing noncommutativity, it is more
convenient to work with Poisson brackets (\ref{V}) than
$\alpha$-star deformed Poisson brackets (\ref{T}). It is important
to note that the relations represented by equations (\ref{T}) are
defined in the spirit of the Moyal product given above. However, in
the relations defined by (\ref{V}), the variables $(x_i, p_j )$ obey
the usual Poisson bracket relations so that the two sets of deformed
and ordinary Poisson brackets represented by relations (\ref{T}) and
(\ref{V}) should be considered as distinct.

In this work, we consider a noncommutative phase space in which
$\theta_{12}=\theta$ and $\beta_{12}=\beta$, such that the
non-vanishing Poisson brackets of the phase space variables are
\begin{equation}\label{W}
\left\{x',y'\right\}=\theta,\hspace{.5cm}\left\{x',p'_x\right\}=\left\{y',p'_y\right\}=1-\theta
\beta/4,\hspace{.5cm}\left\{p'_x,p'_y\right\}=\beta.
\end{equation}
Also, transformations (\ref{U}) takes the form\footnote{It is
important to point out that, transformations (\ref{U1}) can be
inverted if $1-\beta \theta/4\neq 0$.}
\begin{eqnarray}\label{U1}
x'=x-\frac{1}{2}\theta p_y,\hspace{.5cm}p'_x=p_x+\frac{1}{2}\beta y,\nonumber\\
y'=y+\frac{1}{2}\theta p_x,\hspace{.5cm}p'_y=p_y-\frac{1}{2}\beta x.
\end{eqnarray}
On the other hand, to construct Hamilton's equations of motion for
the noncommutative phase space variables, we consider the
Hamiltonian of the noncommutative model as having the same
functional form as equation (\ref{B7}), but in which the dynamical
variables satisfy the deformed Poisson brackets, that is
\begin{equation}\label{X}
{\cal
H}_{nc}=-\frac{1}{2}p'^2_x+\frac{1}{2}p'^2_y+V_0\left(x'-y'\right),
\end{equation}
where by using of (\ref{U1}) takes the form
\begin{equation}\label{X}
{\cal
H}_{nc}=\frac{p^2_y-p^2_x}{2}+\frac{\beta^2}{8}(x^2-y^2)-\frac{\beta}{2}(xp_y+yp_x)-\frac{V_0\theta}{2}(p_x+p_y)
+V_0\left(x-y\right).
\end{equation}
Therefore, the equations of motion read
\begin{equation}\label{Y}
\dot{x}=\left\{x,{\cal H}_{nc}\right\}=-p_x-\frac{\beta}{2}y-\frac{\theta V_0}{2}
,\hspace{.25cm}\dot{p_x}=\left\{p_x,{\cal
H}_{nc}\right\}=-\frac{\beta^2}{4}x+\frac{\beta}{2}p_y-V_0,
\end{equation}
\begin{equation}\label{Z}
\dot{y}=\left\{y,{\cal H}_{nc}\right\}=p_y-\frac{\beta}{2}x-\frac{\theta V_0}{2},\hspace{.25cm}\dot{p_y}=\left\{p_y,{\cal
H}_{nc}\right\}=\frac{\beta^2}{4}y+\frac{\beta}{2}p_x+V_0.
\end{equation}Upon integration of these equations we get the following
expressions for the functions $x(t)$ and  $y(t)$,
\begin{equation}\label{AB}
x(t)=Ae^{\beta t}+Be^{-\beta t}+ht+C_1,
\end{equation}
\begin{equation}\label{AC}
y(t)=-Ae^{\beta t}+Be^{-\beta t}+ht+C_2,
\end{equation}
where $h\equiv(1-\theta \beta/4)V_0/\beta$ and $A$, $B$, $C_1$ and
$C_2$ are integration constants which their values are restricted to
satisfy the constraint equation ${\cal H}_{nc}=0$, that is
\begin{equation}
2\beta^2AB=h(C_1-C_2).
\end{equation}
A calculation similar to what we have done in commutative case
yields
\begin{equation}\label{NL}
N^2(t)=e^{-\phi}/\tilde{a}^2=\frac{1}{4(B(1+\frac{\beta
\theta}{4})e^{-\beta t}+\frac{h^2}{V_0}t+D)},
\end{equation}
\begin{equation}\label{NSF}
a^2(t)=\tilde{a}^2e^{-\phi}=4 A((1+\frac{\beta \theta}{4})e^{\beta
t}+\frac{B\beta^3}{V_0}).
\end{equation}
The form of the scale factor may be simplified if we impose the
initial condition $a(0)=0$, for which we obtain
\begin{equation}\label{NSF}
a^2(t)=4 A(1+\frac{\beta \theta}{4})(e^{\beta t}-1).
\end{equation}
Now, it can be seen that while for the early times of cosmic
evolution the scale factor behaves as $a(t)\propto\sqrt{t}$, i.e.,
is consistent with the behavior of the commutative scale factor, its
late time behavior is like as an accelerated de-Sitter universe,
that is $a(t)\propto e^{\beta t/2}$. This may be interpreted such
that the NC parameter $\beta$ can play the role of cosmological
constant and in this way may be considered as a candidate for dark
energy. This means that the noncommutative effects derive the
universe toward an accelerating phase without the need for any
additional fields. As a matter of fact, we should note that the
noncommutativity in the momentum sector of the phase space (with
noncommutative parameter $\beta$) has its roots in the string theory
corrections to the Einstein gravity \cite{Nima1}. In this sense,
such noncommutative models may be considered as an effective theory
to describe the quantum effects in cosmology. Here, the introduction
of noncommutativity causes additional terms to appear in Hamiltonian
(\ref{X}) as compared to (\ref{B7}) which may be interpreted as the
effects of high energy corrections of a full theory \cite{Nima2}.
Now, since our model can address the issue of dark energy in a
simple cosmological setting, it is reasonable to consider the
noncommutative relations (\ref{W}) as an alternative to the model
theories with dark energy and the accelerating universe. Also, we
see that unlike the commutative case\rlap,\footnote{For $x_0=0$, the
lapse function (\ref{O0}) had a singularity at $t=0$.} in the NC
frame the lapse function (\ref{NL}) is free of singularity. It is
important to note that the NC parameter $\beta$ is responsible for
removing this singularity and from this point of view the role of
the NC parameter $\theta$ is only to shift when the singularity
occurs. To illustrate this issue we may set $\beta=0$ in the
equations (\ref{Y}) and (\ref{Z}) and solve them once again to
obtain
\begin{equation}\label{M1}
x(t)=\frac{1}{2}V_0t^2-(p_{0x}+\theta V_0/2)t+x_0,\hspace{.5cm}p_x(t)=-V_0t+p_{0x},
\end{equation}
\begin{equation}\label{N1}
y(t)=\frac{1}{2}V_0
t^2+(p_{0y}-\theta V_0/2)t+y_0,\hspace{.5cm}p_y(t)=V_0t+p_{0y},
\end{equation}
in which the Hamiltonian constraint impose the following relation
between the constants of integration
\begin{center}$p_{0x}^2-p_{0y}^2-2V_0 (y_0-x_0)+\theta(p_{0x}+p_{0y})V_0=0$.\end{center}
If as the commutative case we choose  $x_0=y_0$, unlike what we saw
in the commutative case this constraint prevents the relation
$p_{0x}=p_{0y}$ between the initial values of the momenta. Indeed,
the constraint leads to $p_{0y}-p_{0x}=\theta V_0$, and hence the
lapse function and the scale factor corresponding to the solutions
(\ref{M1}) and (\ref{N1}) take the forms
\begin{equation}
N^2(t)=\frac{1}{2\left(V_0t^2-\theta V_0 t+2x_0-\frac{\theta^2 V_0}{2}\right)}=\frac{1}{2V_0\left[\left( t-\frac{\theta}{2}\right)^2-\frac{3\theta^2}{4}
\right]+4x_0},
\end{equation}
\begin{equation}
a^2(t)=4|p_{0x}+ \theta V_0/2|(t+\frac{\theta}{2}).
\end{equation}
As functions of time, these have the same time dependence as the
commutative case, i.e. equations (\ref{O0}) and (\ref{O11}), but
with a constant time-shift due to the existence of the NC parameter
$\theta$. This means that the singularity has not removed and in
this sense role of the parameter $\theta$ is only to change the
initial conditions from which the cosmological functions began their
evolution.

\section{Summary}
In this letter we have studied the classical evolution of a modified
$f(R)$ cosmology with a noncommutative phase space. Motivation of
such a study is that introduction of a deformed phase space can be
interpreted as an effective theory which may bring some signals from
a quantum theory. In this work after considering the scale factor
and a scalar field and their momenta as the phase space variable, we
introduced a set of new variables in terms of which the
corresponding minisuperspace takes the form of a Minkowskian space.
In this space we have dealt with the commutation relationships
between variables. In the case where the phase space variables obey
the usual Poisson algebra, i.e., they Poisson-commute with each
other, and for a power law expression for the $f(R)$ function, the
evolution of the universe is like the motion of a particle
(universe) moving on a plane with a constant acceleration. We saw
that in this case the universe decelerates its expansion both in
early and late times of cosmic evolution, in contrast to the current
observation data. We then looked for an accelerated phase of the
universe, suggested by recent supernova observation, in the context
of a deformed phase space of our $f(R)$ model. We showed that a late
time acceleration will occur in the history of the universe due to
introduction of noncommutativity in phase space. We have found that
while the noncommutative parameter between the momenta is
responsible for this acceleration and also removing a singularity
from the classical model, the noncommutative parameter between the
configuration variables only causes a time shift in the commutative
solutions. All in all, our study showed that while the usual
classical model cannot give an accelerated universe, a classical
model with noncommutative phase space variables can derive a late
time acceleration compatible with the current observations. In
particular, we showed that the existence of the deformation $\beta$
parameter in the momentum sector can play the role of a cosmological
constant at late times which means that such a noncommutativity can
be related to the issue of the dark energy. However, note that in
spite of the nature of the cosmological constant in general
relativity as a parameter which is introduced by hand into the
action to fix the problem of the static universe, here its
appearance at late times is a result of some noncommutative
structures which in turn, may be considered as a signal from the
quantum effects in gravity.

\end{document}